\documentclass[10pt]{article}

\usepackage{filecontents}

\RequirePackage[OT1]{fontenc}
\RequirePackage{amsthm,amsmath}
\RequirePackage{natbib}
\RequirePackage[colorlinks,citecolor=blue,urlcolor=blue]{hyperref}

\usepackage{abstract}
\usepackage{mathrsfs}
\usepackage{url}
\usepackage{graphicx}
\usepackage[nottoc]{tocbibind}
\usepackage{fancyhdr}
\usepackage{amsmath}
\usepackage{listings}
\usepackage{lscape}
\usepackage{amssymb}
\usepackage[english]{babel}
\usepackage[autostyle]{csquotes}
\usepackage{setspace}
\usepackage{amsmath}
\usepackage{verbatim}
\usepackage{subcaption}
\usepackage{bbm}
\usepackage[dvipsnames]{xcolor}

\usepackage{rotating}
\usepackage{titlesec}
\usepackage{wasysym}
\usepackage{tikz}
\usepackage{graphicx}
\usepackage{multirow}
\usepackage{amsfonts}
\usepackage{amsmath}
\usepackage{amssymb,graphicx}
\usepackage{pifont}

\graphicspath{ {images/} }

\begin{document}
\newcommand*{\argmin}{\operatornamewithlimits{argmin}\limits}
\newcommand*{\argmax}{\operatornamewithlimits{argmax}\limits}

\def\spacingset#1{\renewcommand{\baselinestretch}%
{#1}\small\normalsize} \spacingset{1}

%%%%%%%%%%%%%%%%%%%%%%%%%%%%%%%%%%%%%%%%%%%%%%%%%%%%%%%%%%%%%%%%%%%%%%%%%%%%%%

{
  \title{\bf {Semiparametric Bayesian Forecasting of Spatial Earthquake Occurrences}}

%\author{Aleksandar A. Kolev$^{1,2}$ \and Gordon J. Ross$^2$ }

\author{Aleksandar A. Kolev$^{1,2}$ \\ email \href{mailto:aleksandar.kolev@ed.ac.uk}{aleksandar.kolev@ed.ac.uk} 
   \and Gordon J. Ross$^2$ \\ email \href{mailto:gordon.ross@ed.ac.uk}{gordon.ross@ed.ac.uk} }

\date{%
\footnotesize
    $^1$Department of Statistical Science, University College London, London, United Kingdom\\%
    $^2$School of Mathematics, University of Edinburgh, Edinburgh, United Kingdom\\% [2ex]%
              \large
}
  \maketitle
} 

\bigskip
\begin{abstract}
Self-exciting Hawkes processes are used to model events which cluster in time and space, and have been widely studied in seismology  under the name of the Epidemic Type Aftershock Sequence (ETAS) model. In the ETAS framework, the occurrence of the mainshock earthquakes in a geographical region is assumed to follow an inhomogeneous spatial point process, and aftershock events are then modelled via a separate triggering kernel. Most previous studies of the ETAS model have relied on point estimates of the model parameters due to the complexity of the likelihood function, and the difficulty in estimating an appropriate mainshock distribution. In order to take estimation uncertainty into account, we instead propose a fully Bayesian formulation of the ETAS model which uses a nonparametric Dirichlet process mixture prior to capture the spatial mainshock process. Direct inference for the resulting model is problematic due to the strong correlation of the parameters for the mainshock and triggering processes, so we instead use an auxiliary latent variable routine to perform efficient inference.

\end{abstract}

\noindent%
{\it Keywords:} Dirichlet Process, ETAS, Spatial analysis, Seismology, Bayesian analysis  
\vfill

\newpage
\spacingset{1} % DON'T change the spacing!
\section{Introduction}
% https://www.tandfonline.com/doi/pdf/10.1198/016214502760046925?needAccess=true
% https://projecteuclid.org/download/pdfview_1/euclid.aoas/1414091236 
% https://projecteuclid.org/download/pdfview_1/euclid.aoas/1419001742
% https://amstat.tandfonline.com/doi/pdf/10.1198/016214503000000710?needAccess=true 
% https://link.springer.com/content/pdf/10.1007%2Fs10463-009-0268-7.pdf

The Hawkes process  \citep{hawkes1971spectra} is a widely used point process model for describing events which are clustered in time and/or space. Applications of the Hawkes process span a diverse range of fields, such as  credit risk \citep{errais2010affine}, criminology \citep{mohler2011self}, neuroscience \citep{chornoboy1988maximum}, social interaction modelling \citep{crane2008robust}, and terrorism \citep{porter2012self}. A particular form of the Hawkes process, known as the Epidemic-Type Aftershock Sequence (ETAS) model, is used in seismology to forecast earthquake occurrences and quantify seismic risk \citep{ogata1988statistical, marzocchi2009real,  schoenberg2013facilitated, omi2014estimating, omi2015intermediate, fox2016spatially}.

The standard ETAS model has two components: an inhomogeneous spatial background process which describes the long-term rate of mainshock activity in a geographical area, and a triggering component which describes the rate and location of aftershock events.  Parameter estimation in the ETAS model is known to be difficult for two reasons \citep{veen2008estimation,zhuang2002stochastic}: first, the likelihood function is  complex and exhibits multi-modality and extended flat regions. Second, the model requires a specification of the inhomogeneous background process which can be challenging when only a limited number of mainshock events are available. Due to these complications, the majority of the ETAS literature focuses only on point estimates of the model parameters and ignores any estimation uncertainty, although some recent exceptions are \citep{fox2016spatially} and \citep{Wang2010}.

While the majority of previous work on the ETAS model is carried out in a frequentist framework, the Bayesian paradigm provides a natural alternative for incorporating parameter uncertainty. However the application of Bayesian methods to the ETAS model have been limited by the complexity of the likelihood function, and many of the approaches which purport to be Bayesian still rely on point estimates of model parameters \citep{ebrahimian2013adaptive} with the prior only used for regularisation purposes. Limited attempts have been made at performing full Bayesian inference for the temporal ETAS model (without a spatial component) using Markov Chain Monte Carlo \citep{omi2015intermediate,Ebrahjmian2017, kumazawa2014nonstationary} but a recent technical report by one of the current authors \citep{Ross2018} casts doubt on the validity of this procedure since the high correlation of parameters in the likelihood function means the effective sample size will be unreasonably low even for a large number of simulations. Additionally, the Bayesian estimation of the inhomogeneous background rate has not to our knowledge been previously considered.

In this paper we present the first fully Bayesian treatment of the ETAS model. Our main innovation is the use of a nonparametric Dirichlet process prior to estimate the inhomogenous spatial background rate. Due to the complexity of the ETAS likelihood function, direct estimation of the resulting nonparametric model is not feasible. As such, we develop a novel estimation procedure using auxiliary latent variables which is an extension of recent work by \citep{Markwick2020} originally proposed for temporal (non-spatial) Hawkes processes. Although our application here is focused on seismology, a similar version of our model could be deployed in other spatial Hawkes applications such as modelling crime hotspots \citep{mohler2011self}.

We begin in Section \ref{section: Intro ETAS} with a brief introduction to the spatial ETAS model and its use in  modelling clustered event sequences. Then in Section \ref{section:KDE/DP} we introduce the Dirichlet process prior that will be used for nonparametric spatial estimation. We next discuss posterior inference strategies using auxiliary latent variables in Section \ref{Section:Bayesian-space}. Issues relating to prior choice and goodness-of-fit testing are discussed in Sections \ref{section:prior_choice_implementation_Details}, \ref{sec:model_comp} and \ref{sec:forecasting}. Finally, we study the performance of the resulting spatial ETAS model on synthetic and real earthquake catalogs in Sections \ref{Section:Application-space} and  \ref{Section: Real_eartquakes} respectively.

\section{Spatio-temporal ETAS model} \label{section: Intro ETAS}

In the ETAS formulation \citep{ogata1998space}, earthquakes are modelled as events $(t_i,m_i,x_i,y_i)$ from a marked point process on the temporal interval  $[0,T]$ and spatial region $\Sigma$, where $t_i$ denotes the time at which the $i^{th}$ earthquake occurred, with $m_i \in \mathbb{R}^+ $ and $(x_i,y_i) \in \Sigma$ denoting its magnitude and  spatial location. It is well known that earthquakes cluster together in both space and time, since large earthquakes tend to trigger further earthquakes nearby, known as aftershocks \citep{utsu1995centenary}. To capture this behaviour, the ETAS point process uses the following conditional intensity function:
\begin{equation} \label{eqn:intensity_space_etas}
   \lambda(t,m,x,y | \mathscr{H}_{t})=\mu(x,y)+ \sum_{t_i<t} \iota(m_i-M_0)  r(t-t_i)  s(x-x_i, y-y_i), 
\end{equation}
where $\mathscr{H}_{t}=\{(t_i, m_i, x_i, y_i); t_i<t\}$  denotes the history of the process before time $t$, and $M_0$ is the magnitude of completeness of the catalog, which is determined empirically and corresponds to the minimum magnitude above which all earthquakes are successfully detected \citep{gutenberg1944frequency, wiemer2000minimum}. The magnitudes are then assumed to be independent and identically distributed according to the usual Gutenberg-Richter (G-R) law $m_i-M_0 \sim Exponential(\beta)$ \citep{gutenberg1944frequency, fox2016spatially}. As a point of notation, we will suppress all explicit dependence on $\mathscr{H}_{t}$ and simply write $\lambda(\cdot)$ for the conditional intensity function.

The function $\mu(\cdot) > 0$ is known as the background process and specifies the baseline intensity, while the functions $\iota(\cdot), r(\cdot), s(\cdot)$  determine the contribution of each previous earthquake at $t_i < t$ to the intensity at time $t$, and are typically chosen to be monotonic decreasing. This implies that each earthquake causes the process intensity to temporarily increase for a period of time, producing local clusters of events.

The ETAS model  can equivalently be viewed as a branching process, as first noted by \citep{hawkes1974cluster}. At each time $t$, suppose that $n_t$ events occurred prior to $t$. From Equation \ref{eqn:intensity_space_etas}, the conditional intensity at time $t$ is a linear superposition of $n_t + 1$ independent inhomogeneous Poisson processes, where the first is the background process which contributes intensity $\mu(x,y)$ and the remainder are indexed by each of the $n_t$ previous events, with each contributing intensity $ \iota(m_i-M_0) \times r(t-t_i) \times s(x-x_i, y-y_i)$ respectively. Since these processes are independent, each event time $t_i$ can be assumed to have been generated either by the background process $\mu(x,y)$ processes triggered by one of the previous events. As a point of convention, we will refer to the events from the background process as immigrant events or mainshocks, with the triggered events being aftershocks.

\begin{figure}[ht!]
\includegraphics[width=0.9\textwidth]{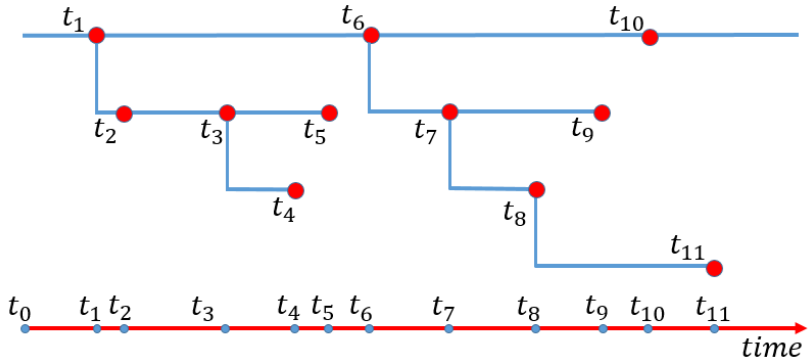}
\centering
\caption{Example of a Branching structure}
\label{fig:branching_updated-space}
\end{figure}

A visual example of such a branching structure is shown in Figure \ref{fig:branching_updated-space}. Events $t_1$, $t_6$ and $t_{10}$ are the immigrants that initiate all other events and which are uncaused by any other event in the sequence. The events $t_2$, $t_3$ and $t_5$ are caused by $t_1$, while $t_4$ is an aftershock of $t_3$. Similarly $t_7$ and $t_9$ are aftershock of $t_6$, while $t_{11}$ is caused by $t_8$, which is a aftershock of $t_7$.  There are no detected aftershock events for $t_{10}$, although some might occur in the future.

\subsection{Specific functional form.}
To complete the specification of the ETAS model, we need to make choices for the functions $\mu(\cdot)$, $r(\cdot)$ $\iota(\cdot)$ and $s(\cdot)$. It is common to take $r(\cdot)$ to be the modified Omori law which has been empirically shown to capture the temporal decay of earthquake productivity \citep{utsu1995centenary}:

\[
r(z)=\frac{K}{(z+c)^p},
\]
where $c$ and $p$ are parameters controlling the decay rate, while $K$ controls the average productivity. Similarly, many empirical studies \citep{ogata1988statistical, ogata2006space, ogata2011significant, fox2016spatially} have shown that a good choice for $\iota(\cdot)$ is:

$$\iota(m_i - M_0) = e^{\alpha(m_i-M_0)}.$$

Several forms have been proposed for the spatial kernel $s(\cdot)$ \citep{ogata2011significant, fox2016spatially}, however we will focus on the most widely used version:

\begin{equation} \label{s_ETAS}
s(x-x_i, y-y_i) =  \Big\{(x-x_i)^2+(y-y_i)^2+d \Big\}^{-q},
\end{equation}
with parameters $d$ and $q$, which corresponds to long-range power law decay.

Finally, the background intensity $\mu(x,y)$ defines the long term average seismicity over the spatial region being considered. This is the most difficult function to choose, since it will depend on the particular spatial region being studied.  Many studies assume that seismicity is constant over space so that $\mu(x,y) = \tilde{\mu}$, however this is highly unrealistic since it is known that earthquakes tend to occur along geological fault lines. As such, several methods have been proposed to model and estimate the spatial dependence of $\mu(x,y)$, such as \citep{ogata2011significant, fox2016spatially,Gerstenberger2004,Helmstetter2006}, which typically uses some kind of kernel smoothing. However the limitation of this approach is that it is difficult to capture the inherent uncertainty in the estimate, which can be especially problematic in regions where only a small number of historical earthquakes are mainshocks from $\mu(x,y)$ rather than triggered events.

The primary goal of this paper is to develop a novel nonparametric Bayesian method for the efficient estimation of $\mu(x,y)$ which captures all underlying uncertainty. We will take this up in Section \ref{section:KDE/DP} after first discussing the  likelihood function for the ETAS model.

\subsection{(Log-)Likelihood}
\label{Sec:loglik}
The log-likelihood function for a general space-time point process with intensity $\lambda(\cdot)$ , observed data $\mathscr{H}_{t}=\{(t_i, m_i, x_i, y_i);$ $t_i<t \}$ and parameter vector $\Theta$ is given by the following expression \citep{daley2003introduction}: 

\begin{multline}
 \label{ETAS_ll_basic}
l(\mathscr{H}_{T}| \Theta)=  \sum_{i=1}^{n}  \log(\lambda(t_i,m_i,x_i,y_i|\Theta)) \\- {\int_{-\infty}^{\infty} \int_{-\infty}^{\infty} \int_{0}^{T} \lambda(z,m,x,y| \Theta) dz dx dy.}
 \end{multline}

The evaluation of the triple integral term in Equation \ref{ETAS_ll_basic} is slow and can be numerically unstable, and so several approximations are provided in the literature \citep{harte2012bias, ogata1998space, schoenberg2013facilitated, lippiello2014parameter}. The calculation of this integral is more feasible if the spatial and temporal kernels $r(\cdot)$ and $s(\cdot)$ are reparameterised to be probability densities that integrate to $1$, which is achieved by splitting $K$ into a product of normalisation constants $K= \bar K K_r K_s$ where:

$$K_r = (p-1)c^{p-1} : \int_0^\infty \frac{K_r}{(z+c)^p} dz = 1,$$
$$K_s = \frac{q-1}{\pi d^{1-q}} : \int_{-\infty}^{\infty} \int_{-\infty}^{\infty} K_s (x^2+y^2+d)^{-q} dx dy = 1.$$

Then 
$$r'(z)=  \frac{\bar{K}r(z)}{K} \quad \text{and} \quad s'(z)= K_s s(z).$$

Using similar notation, we write $\mu(x,y) = \tilde{\mu} \phi(x,y)$ where $\phi(\cdot)$ is a probability density that integrates to 1.
Under this notation, the triple integral in Equation \ref{ETAS_ll_basic} can then be approximated as:
\[
\int_{-\infty}^\infty \int_{-\infty}^\infty \int_0^\infty \lambda(z_i,m_i,x_i,y_i| \Theta) dz dx dy    \approx \tilde{\mu} T + \bar K \sum_{i=1}^{n} e^{\alpha(m_i-M_0)}.
\]

This is only an approximation since it assumes that the temporal and spatial domains are infinite, when in fact the point process is defined on the finite region $[0,T]\times \Sigma$. In \cite{schoenberg2013facilitated}, it was shown that the assumption of an infinite spatial domain has negligible effect on the likelihood and so we will use this approximation to reduce the computational cost of evaluating the triple integral. However, the assumption of an infinite time domain can be problematic. In \cite{kolev2019inference}, it was shown that the infinite time assumption provides a poor performance regarding the North California seismic sequence, thus we prefer not to use it whenever possible. As such, we will avoid using it and instead work on the finite temporal region $[0,T]$.

In summary, the log-likelihood of the Spatial ETAS model based on the finite time and infinite space assumptions can be approximated by the following expression: 

\begin{equation} \label{eqn:space_ETAS_ll_approx}
\begin{split}
l(\mathscr{H}_{T} | \Theta)=  \sum_{i=1}^{n}  log(\lambda(t_i,m_i,& x_i,y_i| \Theta)) - \tilde{\mu} T \\&+ \bar K \sum_{i=1}^{n} e^{\alpha(m_i-M_0)}\Bigg( 1- \frac{c^{p-1}}{(T-t_i+c)^{p-1}} \Bigg), 
\end{split}
\end{equation}
with a parameter vector $\Theta=(\tilde{\mu}, \alpha, c, p, \bar{K}, d, q).$ along with the parameters that definite the background kernel $\phi(x,y)$.

\section{Nonparametric Estimation of Background Intensity}
\label{section:KDE/DP}
We will consider a variety of methods for estimating the spatial background rate $\mu(x,y)$ in the ETAS model. Using the parametrisation above, we can write:

$$\mu(x,y) = \tilde{\mu} \phi(x,y),$$
where $\tilde{\mu}$ is a scaling constant and $\phi(x,y)$ is a probability density that integrates to $1$. We will consider three different Bayesian models for the background rate. In all three, the constant $\tilde{\mu}$ will be estimated separately, with the models varying in terms of how they treat $\phi(x,y)$:

\begin{enumerate}
    \item $\phi(x,y) \propto 1$, in which case the background intensity is constant over space. This is highly unrealistic due to the known fact that seismic activity is highly spatially dependent, and we use this model only as a baseline.
    
    \item A nonparametric model where $\phi(x,y)$ is learned using Kernel Density Estimation (KDE). Different versions of this method are fairly common in the seismology literature, albeit in a non-Bayesian context \citep{zhuang2002stochastic, marsan2008extending, sornette2009limits, fox2016spatially,Helmstetter2006}.  Note that fitting a KDE directly to the observed earthquake catalogs will result in a biased estimate of $\phi(x,y)$ since $\phi(x,y)$ is specifically a model for the background (immigrant) events only, and not for the triggered events. 
    
    \item A nonparametric model where $\phi(x,y)$ is learned in a fully Bayesian manner using a spatial Dirichlet Process prior, in a way which distinguishes between background and triggered events. This is substantially more complex than the KDE approach since it require declustering the earthquakes into background and immigrant events, with only the background events used to estimate $\phi(x,y)$. This corrects the bias in the KDE approach.
\end{enumerate}

The first model using the uniform density is self-explanatory. We will now discuss the other two in more detail.
 
\subsection{KDE ETAS} \label{Sec: KDE ETAS}
The second method described above uses kernel density estimation to learn $\phi(x,y)$. Suppose we observe $n$ observations $(t_i, m_i, x_i, y_i)$ from the point process. Let $\mathbf{z}_i = (x_i,y_i)$ be the spatial coordinates of the $i^{th}$ earthquake. Then a KDE estimate of $\phi(\cdot)$ can be given by:

$$\hat{\phi}(\mathbf{z})={\frac {1}{n}}\sum _{i=1}^{n}K_{\mathbf {H} }(\mathbf{z} -\mathbf {z} _{i})$$
with 
$$K_{\mathbf {H} }(\mathbf {x} )=|\mathbf {H} |^{-1/2}K(\mathbf {H} ^{-1/2}\mathbf {x} ),
$$
where $\mathbf{H}$ is $d \times d$, symmetric and positive-definite which is also referred to as bandwidth matrix and $K(\cdot)$ is a symmetric kernel function. Without loss of generality we can choose it to be: 
\[ K(\mathbf {x} )= (2 \pi)^{-1} \exp(-\frac{1}{2}\mathbf{x}'\mathbf{x}).\]

The estimated  $\phi(x,y)$ is then treated as being a fixed constant while the other model parameters including $\tilde{\mu}$ are estimated. We refer to this model as KDE ETAS. Although Kernel Density Estimation is a powerful and flexible nonparametric method, it has two drawbacks \citep{marsan2008extending, sornette2009limits}. First, it can be difficult to choose $\mathbf {H}$ in a manner which produces an accepted level of smoothing across the whole spatial domain rather than under/over fitting in particular regions. Second, the resulting KDE estimate of $\phi(\cdot)$ is based on smoothing over all $n$ earthquakes in the historical catalog. However this is not the correct behaviour in the context of an ETAS model, since the $\phi(\cdot)$ function specifies the occurrence of only background, rather than triggered events. As such, we would expect the KDE estimate to be biased, and assign too much probability mass to spatial regions where large magnitude earthquakes have occurred, since their large number of triggered aftershocks will be incorporated into the estimate of $\phi(\cdot)$.  To some extent this problem can be avoided by first declustering the catalog before fitting the KDE as in \cite{Helmstetter2006} however it is difficult to choose an adequate bandwidth when fitting to such a declustering catalog, and the inability to incorporate estimation uncertainty into the KDE is problematic when fitting to declustered catalogs where the number of background events may be low. These problems are avoided in the fully Bayesian approach which we will now discuss.

\subsection{DP ETAS} \label{Sec: DP ETAS}

To avoid the drawbacks of KDE, we propose estimating $\phi(x,y)$ using a Dirichlet process mixture prior. The Dirichlet process (DP) was introduced by  \cite{ferguson1973bayesian, antoniak1974mixtures} as a probability distribution over probability distributions, and is commonly used as a prior in Bayesian nonparametric modelling. If a probability distribution $G$ has a DP prior then we write:

\[G \sim DP(\chi, G_0),\] 
where  $G_0$ is the base distribution which defines the expected value of the DP and $\chi>0$ is a measure of the variance. The DP is a conjugate prior in the following sense: suppose that $\theta_1,\ldots, \theta_n \sim G$ where $G\sim DP(\chi,G_0)$. Then the posterior distribution of $G$ is:

\[G|\theta_1, \cdot \cdot \cdot, \theta_n \sim DP\Bigg(\chi+n, \frac{\chi G_0+\sum_{i=1}^n \delta_{\theta_i}}{\chi+n}\Bigg),\] 
where $\delta_{\cdot}$ is a Dirac delta function  \citep{dirac1947principles}.

A constructive definition of the DP was given by  \cite{sethuraman1994constructive}, who showed that samples from a DP can be written in stick breaking form:
\[G=\sum_{i=1}^{\infty} \pi_i \delta_{\psi_i},\quad \psi_k \sim G_0,\]
where $\{\beta_i\}_{i=1}^{\infty} \sim Beta(1, \chi)$,   $\pi_k=\beta_k\prod_{i=1}^{k-1}(1-\beta_i)$, and $ \delta_{\psi_k}$ is the Dirac delta function \citep{dirac1947principles}. This provides a practical method for drawing a sample from a DP, by approximating the stick breaking as a finite (truncated) sum:
\[G=\sum_{i=1}^{N} \pi_i \delta_{\psi_i},\quad \psi_k \sim G_0.\] Combining this with the conjugacy result above, we can hence sample $G$ from its posterior distribution given some observed data $\theta_1,\ldots,\theta_n$ as:

\[G |\theta_1, \cdot \cdot \cdot, \theta_n=\sum_{i=1}^{N} \pi_i \delta_{\psi_i},\quad \psi_k \sim \frac{\chi G_0 + \sum_{i=1}^n \delta_{\theta_i}}{\chi+n},\]
where 
$\{\beta_i\}_{i=1}^{N} \sim Beta(1, \chi+n)$ and $ \pi_k=\beta_k\prod_{i=1}^{k-1}(1-\beta_i).$

An alternative representation of the DP is based on the Chinese restaurant process \citep{neal2000markov}, which shows that the marginal prior distribution of the samples $\theta_1,\ldots,\theta_n$  (with $G$ integrated out) can be written as:
\[\theta_i | \theta_1, ..., \theta_{i-1} \sim \frac{1}{i-1+\chi} \sum_{j=1}^{i-1} \delta_{\theta_i} + \frac{\chi}{i-1+\chi} G_0,\]
where $\theta_i \sim G_0$

\paragraph{The Dirichlet Process as a Spatial ETAS Prior}
\label{sec:DP_etas_intro}
In this work, we propose to use the Dirichlet Process (DP) as a nonparametric prior for the background ETAS intensity $\phi(x,y)$. From the above results, we can see that samples from a DP follow a discrete distribution. In order to adapt the DP to continuous data, it is common to instead use it as a prior distribution for a mixture model. This leads to the following specification:

\[\phi(x,y)= \int k(x,y | \theta) dG(\theta) \]
\[ G \sim DP(\chi, G_0) \]
where $k(\cdot)$ is a mixture kernel. This formulation corresponds to an infinite dimensional mixture model where the DP is used as a prior on the mixing distribution parameter.

Since $\phi(x,y)$ is a two-dimensional spatial distribution, we will model it as a mixture of bivariate Gaussians, where $\theta = (\boldsymbol{\mu}, S)$ is the mean vector and precision matrix. For conjugacy, we choose $G_0$ to be the Normal Inverse-Wishart distribution. This leads to the following model:

\[x_i,y_i|c_i \sim N(\boldsymbol{\mu_i}, (S_{i})^{-1}) \]
\[\boldsymbol{\mu_i}, S_i \sim G\]
\[G \sim DP\Big(\chi, NW (\boldsymbol{\xi}, \rho, \beta, \beta V)\Big)
\]
where $\boldsymbol{\xi}, \rho, \beta$ and $V$ are the parameters of the Normal Inverse-Wishart distribution where the mean $\boldsymbol{\mu_k}$ follow a Gaussian distribution $ \boldsymbol{\mu_k} \mid S_k, \boldsymbol{\xi}, \rho \sim N(\boldsymbol{\xi}, (\rho S_k)^{-1})$ and the precision matrix $S_k$ comes from a Wishart distribution \citep{wishart1928generalised}
$ S_k \mid \beta, V \sim  W(\beta, (\beta V)^{-1}). $

From here after, this ETAS alternative will be referred to as DP ETAS model.

\section{Posterior Simulation} \label{Section:Bayesian-space}
We propose to estimate all model parameters $\Theta = \{\tilde{\mu},\phi(x,y), \alpha, \bar{K}, c, p, d, q\}$ of the ETAS model using Bayesian inference. Given a prior distribution $p(\Theta)$ and earthquake catalog data $\mathscr{H}_{T}$, the resulting posterior is:

\begin{equation}
\label{eqn:Bayesian_ration-space}
\pi(\Theta|\mathscr{H}_T ) = \frac{p(\mathscr{H}_T| \Theta) \pi(\Theta)}{\int_\Theta{p(\mathscr{H}_T| \Theta) \pi(\Theta) d \Theta} }.
\end{equation}

The multi-dimensional integral in Equation \ref{eqn:Bayesian_ration-space} cannot be solved analytically for the ETAS model. For this reason, we will use simulation methods to approximate the posterior instead.

It may initially seem feasible to use  Metropolis-Hastings or Gibbs sampling to draw samples from the posterior. However this are serious grounds to doubt that this will work. Recall from Equation \ref{eqn:space_ETAS_ll_approx}, that the log-likelihood  of the ETAS model is:

\begin{equation} 
\begin{split}
& l(\mathscr{H}_{T}  | \Theta) \approx  - \tilde{\mu} T + \bar K \sum_{i=1}^{n} e^{\alpha(m_i-M_0)}\Bigg( 1- \frac{c^{p-1}}{(T-t_i+c)^{p-1}} \Bigg)
\\& +\sum_{j=1}^{n}  log\Big(\mu(x_j,y_j)+ \sum_{t_i<t_j} \iota(m_i-M_0) \times r(t_j-t_i) \times s(x_j-x_i, y_j-y_i)\Big) , 
\end{split}
\end{equation}

The first problem is that evaluating the likelihood function requires a double summation and this  evaluation must take place each time a new parameter value is proposed. As such, direct MCMC would be computationally very demanding, and cannot feasibly be run on a catalog containing more than a few hundred earthquakes. The second problem is that \cite{schoenberg2013facilitated} studied the performance of frequentist maximum likelihood estimation for the ETAS model based on directly maximising the above likelihood function when $\mu(x,y) = \tilde{\mu}$ was a constant value. They found that the resulting parameter estimates often differed substantially from their true values. This is because the likelihood function is multi-modal and the components of the parameter vector are highly correlated. Since MCMC methods can also suffer from serious convergence issues when the parameters are correlated, it is reasonable to believe that this direct MCMC procedure will suffer from the same problem. Since this problem is already present in the simple parametric case with constant $\tilde{\mu}$, it will be even worse in the more complex nonparametric setting.

 As such, we instead propose a reparametrisation of the model based on latent variables that aims to break the parameter correlation in the likelihood function and lead to an efficient Metropolis-Hastings algorithm for posterior sampling. This is an extension of the method proposed by \cite{Markwick2020} for the temporal Hawkes process.

\subsection{Latent Variable Formulation} \label{sec:latent_ETAS_space}

%We now develop an posterior sampling scheme based on introducing latent variables. These have the effect of breaking the dependence between the parameters in the likelihood function. We will show that conditional on the latent variables, the parameter sets $\{\tilde{\mu},\phi(x,y)\}$, $\{\bar{K},\alpha,p,c\}$, and $\{d,q\}$ are all independent of each other, which greatly improves the convergence of MCMC sampling.

As discussed in Section \ref{section: Intro ETAS}, the ETAS model can be reinterpreted as a branching process in the following sense. Suppose that at time $t$ there have been $n_t$ previous events. Then, Equation \ref{eqn:intensity_space_etas} can be interpreted as showing that the intensity function at time $t$ is a sum of $n_t + 1$ different Poisson processes. The first is a time-homogenous Poisson process with intensity $\mu(x,y)$, while each of the other $n_t$ processes is triggered by the previous events. Specifically, for each $1 \leq j \leq n_t$, the event that occurred at time $t_j$ triggers an inhomogeneous Poisson process with intensity:

\begin{equation}
\lambda_j(t,x,y) = Ke^{\alpha(m_j-M_0) } (t-t_j+c)^{-p}\Big\{(x-x_j)^2+(y-y_j)^2+d \Big\}^{-q}.
\label{eqn:kernel}
\end{equation} 

Now consider an event that occurs at time $t_i$, where there have been $n_{t_i}$ previous events. Based on standard results about the superposition of Poisson processes \citep{daley2003introduction} we can interpret event $t_i$ as having been generated by a single one of these $n_{t_i}+1$ processes. For an earthquake catalog that contains $n$ events in total, we hence introduce the latent branching variables $B=\{B_1,\ldots,B_n\}$ where $B_i \in \{0,1,\ldots,n_{t_i}\}$ indexes the process which generated $t_i$:

$$
B_i \sim \left\{ \begin{array}{rl}
0 & \mbox{ if $t_i$ was produced by the background process}\\
 j & \mbox{ if $t_i$ was triggered by the previous earthquake at time $t_j$} \\
       \end{array} \right.
$$

Conditional on knowing $B$, we can partition the earthquakes into $n+1$ sets $S_0,\ldots,S_n$ where:

$$S_j = \{t_i ; B_i = j \}, \quad 0 \leq j  < n,$$
so that $S_0$ is the set of immigrant events which were not  triggered by previous earthquakes, and $S_j$ is the set of direct aftershocks triggered by the earthquake at time $t_j$. It is clear that these sets are mutually exclusive and that their union contains all the earthquakes in the catalog.  Additionally, we can see that the earthquakes in set $S_0$ are generated by an inhomogeneous Poisson process with intensity $\mu(x,y)$, while the events in each set $S_j$ for $j>0$ are generated by a single inhomogeneous Poisson process with intensity given by Equation \ref{eqn:kernel}. 
The ETAS likelihood function from Equation \ref{eqn:space_ETAS_ll_approx} can hence be rewritten (conditional on knowing the latent branching variables) as:

\begin{equation}
\label{eqn:LL_branched-space}
\begin{split}
p(\mathscr{H}_t| \Theta, B)=& e^{-\tilde{\mu} T} \prod_{t_i \in S_0}  \mu(x_i,y_i) \\
&\prod_{j=1}^{n} \Bigg( e^{-  \bar K  e^{\alpha(m_j-M_0)}  \Big( 1- \frac{c^{p-1}}{(t_n-t_i+c)^{p-1}} \Big) } \{\bar K  e^{\alpha(m_j-M_0)}\}^{|S_j|} \Bigg)\\
& \prod_{j=1}^{n}  \prod_{t_i \in S_j} \Bigg( \frac{K_t K_r}{(t_i-t_j+c)^p}((x_i-x_j)^2+(y_i-y_j)^2+d)^{-q}\Bigg),
\end{split}
\end{equation}
where as before $\mu(x,y) = \tilde{\mu} \phi(i,y)$ where $\tilde{\mu}$ is a constant and $\phi(x,y)$ has a KDE or DP mixture prior. From this factorisation, we can see that $\tilde{\mu}$ and $\phi(x,y)$ are independent of the other model parameters in the likelihood, and hence will be independent in the posterior assuming prior independence. This latent variable formulation hence breaks the dependency which would have made the KDE or DP part of the model difficult to learn, and further weakens the dependence between $\{c, p\}$, $\{\alpha, \bar{K}\}$ and $\{d, q\}$ which will allow for more efficient MCMC sampling.

We hence propose a Gibbs sampler which samples the parameters in the following conditionally independent blocks:

\begin{itemize}
    \item $p(B | \tilde{\mu}, \phi(x,y), \bar{K}, \alpha, c, p, d, q,\mathscr{H}_t)$
    \item  $p(\phi(x,y) | B, \mathscr{H}_t)$
    \item  $p(\tilde{\mu} | B, \phi(x,y), \mathscr{H}_t)$
    \item  $p(\bar{K}, \alpha | B,  c, p, \mathscr{H}_t)$
    \item $p(c,p | B, \bar{K}, \alpha, \mathscr{H}_t)$
    \item  $p(d, q| B, \mathscr{H}_t)$
\end{itemize}
     
Given a set of model parameters $\Theta^{(k)}$ at iteration $k$ of the Gibbs sampler, we now explain how to sample the next value $\Theta^{(k+1)}$ from the above full conditional distributions.

\subsubsection{Sampling $B$}
 As shown by \cite{zhuang2002stochastic} in a very different context (stochastic declustering), each individual branching 
variable $B_i^{(k+1)}$  can be sampled exactly from its conditional posterior. Note that each $B_i$ can take values only in the discrete set $\{0,1,\ldots,i-1\}$, i.e. each earthquake can be triggered only by either a previous earthquake, or the background process. Assuming a uniform prior on each $B_i$, the probability of it being caused by any of the $i$ processes is simply the proportion of the overall intensity that can be attributed to that process, i.e.:
%I HAVE DROPPED ALL SUBSCRIPTS HERE!!!!

\begin{equation}
    \label{eqn:branchingposterior}
p(B_i^{(k+1)}=j|\mathscr{H}_t,\Theta^{(k)})=
\begin{cases}
\begin{split}
  & \frac{\tilde{\mu} \phi(x_i,y_i)}{ \lambda(t_i,m_i,x_i,y_i|\mathscr{H}_{t_i}, \Theta^{(k)})} &\mbox{ for } j=0 \\
  & \frac{g(t_i-t_j) h(x_i-x_j, y_i-y_j) \iota(m_i)   }{\lambda(t_i,m_i,x_i,y_i|\mathscr{H}_{t_i}, \Theta^{(k)})} &\mbox{ for } j\neq0\\
\end{split}
 \end{cases}\\
\end{equation}

Each $B_i$ can hence be drawn independently from the discrete distribution on $\{0,\ldots,i-1\}$, with weights given by the above \citep{kolev2019inference}. Note that unlike the MCMC sampler used in \cite{rasmussen2013bayesian} for Hawkes processes, this  samples $B^{(k+1)}$ exactly from its conditional posterior, which drastically improves computational efficiency.%Note that although sampling $B$ is an $O(n^2)$ operation, it only needs to be performed once during each iteration of the sampler and then the other parameters can be efficiently sampled from their conditional posterior. This is the key reason why our scheme is more efficient than the direct approach.

\subsubsection{Update $\phi(x,y)$}
This step is required only for DP ETAS since $\phi(\cdot)$ is constant in the KDE version of the model. Recall from Section \ref{sec:DP_etas_intro} that

$$\phi(x,y) = \int N(x,y | \theta) d\theta$$
$$\theta \sim G$$
$$G \sim DP(\chi, G_0),$$
where $\phi(x,y)$ is the generating function for the $|S_0|$ immigrant earthquakes that are assigned to the background process based on the current branching structure.

In order to simulate a value of $\phi(\cdot)$ from its conditional posterior, we first simulate values of $\theta_i, i \in \{1,2,\ldots,|S_0|\}$ from their posterior distributions given the earthquakes which are assigned to the background process, using the usual Chinese Restaurant process sampler. Given these values, we then have from Section \ref{Sec: DP ETAS} that:

$$G | \theta_1,\ldots,\theta_{|S_0|} \sim DP\Bigg(\chi+n, \frac{\chi G_0+\sum_{i=1}^n \delta_{\theta_i}}{\chi+n}\Bigg).$$

We can hence sample a value of $G$ from its posterior using truncated stick breaking, i.e:

$$G=\sum_{i=1}^{N} \pi_i \delta_{\psi_i}.$$
This hence fully defines a realisation of $\phi(x,y)$ from its posterior. Note that the reason why we need to simulate a realisation of $\phi(x,y)$ (and hence $G$) rather than working only with the $\theta_i$ samples is that we need to have a realisation of $\phi(x,y)$ to evaluate the branching posterior in Equation \ref{eqn:branchingposterior}.

As part of this step, we can also perform an update of the DP hyperparameters of $G_0$ and of $\chi$, if we wish to work with the full hierarchical version of the DP. This can be done by assigning them a sensible prior distribution, such as $\chi \sim Gamma(1,1)$. The updates in this case are standard in the DP literature and are described in detail in \citep{gorur2010dirichlet}.

\subsubsection{Update the value of $\tilde{\mu}$}
 Using Equation \ref{eqn:LL_branched-space}  we can observe that $\tilde{\mu}$ depends only on the number of events in the background process $S_0$, hence:

\begin{equation*}
    \begin{split}
        p(\tilde{\mu}|\mathscr{H}_t, \Theta, B) \propto & \pi(\tilde{\mu})e^{-\tilde{\mu} T} \prod_{t_i \in S_0}  \mu(x_i,y_i)= \\&\pi(\tilde{\mu})e^{-\tilde{\mu} T} \tilde{\mu}^{|S_0|} \prod_{t_i \in S_0}  \phi(x_i,y_i) \propto \pi(\tilde{\mu})e^{-\tilde{\mu} T} \tilde{\mu}^{|S_0|}.
    \end{split}
\end{equation*}

This is equivalent to estimating the intensity $\tilde{\mu}$ of a homogeneous Poisson process on $[0, T]$, with event times $S_0$. In this case, the Gamma distribution is the conjugate prior: $\pi_{\tilde{\mu}}= Ga(\alpha_{\tilde{\mu}}, \beta_{\tilde{\mu}})$. The posterior distribution is then $p(\tilde{\mu}|\mathscr{H}_t, \Theta, B) = Ga (\alpha_{\tilde{\mu}} + |S_0|, \beta_{\tilde{\mu}} + T)$ which can be sampled from directly \citep{Ross2018}.

\subsubsection{Update the values of $\bar{K}$ and $\alpha$}
Similar to the process for sampling $\tilde{\mu}$, we can sample new values of $\bar{K}$ and $\alpha$ from $p(\bar{K}, \alpha|\mathscr{H}_t, \Theta, B)$. Based on Equation \ref{eqn:LL_branched-space}, we conclude that:

\begin{equation*}
    \begin{split}
    p(\alpha, \bar K |\mathscr{H}_t,  \Theta, B) \propto & \pi(\bar K, \alpha) \\ & \prod_{j=1}^{n} \Big( e^{-  \bar K  e^{ \alpha(m_j-M_0)}  \Big( 1- \frac{c^{p-1}}{(t_n-t_i+c)^{p-1}} \Big)  } \{\bar K  e^{\alpha(m_j-M_0)}\}^{|S_j|}\Big). 
        \end{split}
\end{equation*}

Although there is no conjugate prior in this case, it is straightforward to use (e.g.) random walk MCMC to draw a sample from this posterior as described in the beginning of this section.  

\subsubsection{Update the values of $c$ and $p$}
Again, based on Equation \ref{eqn:LL_branched-space}, we can see that the posterior distribution of $c$ and $p$ is given by:

\begin{equation*}
    \begin{split}
        p(c,p|\mathscr{H}_t, \Theta, B) \propto &\pi(c,p) \\ & \prod_{j=1}^{n} \Bigg( e^{-  \bar K  e^{ \alpha(m_j-M_0)}  \Big( 1- \frac{c^{p-1}}{(t_n-t_i+c)^{p-1}} \Big)  }  \prod_{t_i \in S_j} \frac{K_t }{(t_i-t_j+c)^p}\Bigg).
    \end{split}
\end{equation*}

%\[p(c,p|\mathscr{H}_t, \Theta, B) \propto \pi(c,p) \\ \prod_{j=1}^{n} \Bigg( e^{-  \bar K  e^{ \alpha(m_j-M_0)}  \Big( 1- \frac{c^{p-1}}{(t_n-t_i+c)^{p-1}} \Big)  }  \prod_{t_i \in S_j} \frac{K_t }{(t_i-t_j+c)^p}\Bigg).\]

The parameter sampling can be done using (e.g.) standard random walk MCMC sampler.

\subsubsection{Update the values of $d$ and $q$}
As a last step of our MCMC sampler we update the offspring kernel space parameters $d$ and $q$. The expression below is a simplified approximation that depends on an infinite space approximation which was discussed in Section \ref{Sec:loglik}
\[p(d,q|\mathscr{H}_t, \Theta, B) \propto \pi(d,q) \prod_{j=1}^{n} \Big(  \prod_{t_i \in S_j} K_r\Big((x_i-x_j)^2+(y_i-y_j)^2+d\Big)^{-q}\Big).\]
%However, the above result incorporates infinite space assumptions. An alternative that incorporates an additional term inducing dependence between the spatial parameters $d$ and $q$ and the temporal parameters $c$ and $p$, as well as the marginal ones - $\alpha$ and $\bar K$ - can be introduced. However, this is beyond the scope of this paper. Further complications related to the integration of the  $h(\cdot)$ for finite space are also present. Since the offspring detection range is not restricted, we assume that the infinite space assumption holds true. 

\section{Prior Choice, and Implementation Details} \label{section:prior_choice_implementation_Details}

A common problem with the estimation and simulation of an ETAS model is that certain parameter values can result in infinitely many earthquakes being generated from the process with non-zero probability. It can be shown \citep{helmstetter2002subcritical} that to guarantee a finite catalog, the average number of aftershocks produced by each earthquake in the catalog must be less than 1. The intuition for this result is that if the average number of aftershocks is greater than 1, then the branching process representation of the process may never converge.

By integrating over the Gutenberg-Richter magnitude distribution, it can be shown \citep{Ross2018} that the average number of aftershocks will be less than 1 if:

\[\beta<\alpha\]
and 
\[\frac{\bar K \beta}{\beta-\alpha}<1,\]
where $\beta$ is the parameter in the G-R distribution $p(m-M_0|\beta)$ (Section \ref{section: Intro ETAS}). 

As such, we will choose our prior distributions so that positive mass is only assigned to regions where the above parameter relations are satisfied. We choose to use relatively uninformative Uniform priors:  $\alpha \in (0,10)$, $c \in (0,10)$, $p \in (1,30)$, $\bar K \in (0,30)$, $d \in (0, \infty)$ and $q \in (1, \infty)$ , with the regions not satisfying the above relations assigned zero mass. Note that the priors for $c$ and $p$ are slightly informative which is required since these parameters are only weakly identifiable \citep{holschneider2012bayesian}

Finally in the above discussion of the MCMC sampler, we mentioned that random walk Metropolis-Hastings was used to update some blocks of parameters. For these, we used a Normal proposal distribution with standard deviation of 0.1.

\section{Model comparison}  \label{sec:model_comp}
We have proposed three different versions of the Bayesian ETAS model, which respectively use the Uniform distribution, KDE and a DP mixture model to represent $\phi(x,y)$. In this section we discuss the in- and out-of-sample performance metrics that we will use to compare these models. 

\subsection{Deviance Information Criterion (DIC)} \label{sec:DIC}
The DIC is a fully Bayesian alternative to the Akaike Information Criteria (AIC) \citep{akaike1973maximum}. It replaces the maximum likelihood parameter estimates of $\Theta$ with their posterior mean $\bar{\Theta}$. The correction associated with the number of used parameters is substituted with a measure of parameter adequacy ($p_{DIC}$) %based on the goodness of the parameter chain in terms of log-likelihood
\citep{gelman2014bayesian}. Given a set of model parameters $\Theta$ the model's DIC value is:

$$\text{DIC}(\Theta)= - 2l(\mathscr{H}_t|\bar{\Theta})+ 2.p_{DIC} ,$$
where $l(\mathscr{H}_t|\Theta)=log(p( \mathscr{H}_t| \Theta ))$ is the log-likelihood function and $p_{DIC}$ is the effective sample size, which evaluates the number of independent samples the
MCMC draws are equivalent to.  It is defined as: 
$$p_{DIC}= 2l(\mathscr{H}_t|\bar{\Theta}) - 2  \mathop{{}\mathbb{E}}(l(\mathscr{H}_t|\Theta) \cong 2l(\mathscr{H}_t|\bar{\Theta}) - 2 \frac{1}{S} \sum_{s=1}^{S} l(\mathscr{H}_t|\Theta_s),  $$
where $\Theta_s$ indicates the $s$th parameters' sample in the considered MCMC chain. Alternatively, we can compute the effective sample size as the variance of the obtained log-likelihood values for all sampled parameters as follows: 

$$p_{DICalt}= 2Var(l(\mathscr{H}_t|\Theta). $$
This method is not as numerically stable as the other one but it is easier to compute because it does not require the allocation of $\bar{\Theta}$, which is a computationally demanding task with respect to the $\phi(x,y)$ of DP ETAS model. This measure further guarantees to provide positive values. For all these reasons we will use this alternative of the DIC metric through this paper. 

\subsection{Out-of-sample log-likelihood}\label{section:out_of_sample}

A common way to evaluate the performance of earthquake models is to consider the out-of-sample predictive distributions, i.e. how well we can predict the occurrence time and locations of earthquakes in the time window $[T,U]$ given that we have fitted a model to the time window $[0,T]$. Several versions of this approach have been used in the literature, with a summary given in \cite{bray2013assessment}.

Since all of our models are Bayesian with completely specified probability distributions, we will compare based on the out-of-sample posterior predictive likelihood (i.e. the likelihood of the future observations averaged over the samples which have been drawn from the posterior).

\section{Forecasting}
\label{sec:forecasting}

We note that since our models  are fully specified, we can use them to create forecasts about future earthquakes, e.g. the probability of an earthquake with magnitude greater than $M_0$ occurring during some future time period $\mathscr{H}_{new}$ . This can  be done by simulation, based on the samples drawn from the posterior using MCMC to produce simulated point process trajectories over the time period of interest, and then extracting the quantities to be forecasted as summary statistics. This is essentially a Monte Carlo approximation to the forecasting distribution:

$$p(\mathscr{H}_{new} | \mathscr{H}_t) = \int p(\mathscr{H}_{new}|\Theta)p(\Theta| \mathscr{H}_t) d \Theta \approx \frac{1}{M}\sum_{i=1}^M p(\mathscr{H}_{new}|\Theta^{(i)})$$ 
$$\text{for} \quad \Theta^{(i)} \sim p(\Theta| \mathscr{H}_t).$$

\section{Simulation Study} \label{Section:Application-space}
In this section we will use synthetic (simulated) data to evaluate and compare the performance of the three Bayesian ETAS models for $\phi(x,y)$ using 1) the Uniform distribution, 2) KDE, and 3) a Dirichlet process mixture. The next section will similarly compare them using real earthquake catalogs.

\subsection{Initial Comparison} \label{section:intro_f1_f2_f3}
We simulate three catalogs, each with a different choice for the density $\phi(x,y)$. The first density that we consider follows a standard bivariate normal distribution i.e.

\begin{equation} \label{eqn:phi_1}
  \phi_1(x,y) \sim N\Bigg(\begin{bmatrix} 
0  \\
0 
\end{bmatrix}
, \begin{bmatrix} 
1 & 0  \\
0 & 1
\end{bmatrix}\Bigg).  
\end{equation}

The second one is a mixture of two Normal distributions. The first of them has a mean of $(-1, -1)$ and the second one $(1,1)$. They share a common covariance matrix that comprises zero covariance and $0.4$ standard deviation in each dimension i.e. 

\begin{equation} \label{eqn:phi_2}
\phi_2(x,y) \sim N\Bigg(\begin{bmatrix} 
-1  \\
-1 
\end{bmatrix}
, \begin{bmatrix} 
0.4 & 0  \\
0 & 0.4
\end{bmatrix}\Bigg) +
N\Bigg(\begin{bmatrix} 
1  \\
1
\end{bmatrix}
, \begin{bmatrix} 
0.4 & 0  \\
0 & 0.4
\end{bmatrix}\Bigg).
\end{equation}
The third density aims to simulate a seismic fault - all events are uniformly distributed on a line with known boundary conditions. This requires the specification of a fixed spatial region $\Sigma$. We sample uniformly a realisation of the $x$ range of $\Sigma$. Then we transform it into a point on a line defined by an intercept $a$ and a slope $b$ and further scale it by an error component $\epsilon \sim N (0, \sigma_{\epsilon}^2)$ i.e. 

\begin{equation} \label{eqn:phi_3}
\phi_3(x,y)=\phi_x(x)\phi_y(y),
\end{equation}
for $\phi_x(x) \sim Unif(\Sigma_x)$, where $\Sigma_x$ is the range in $x$ dimension and $\phi_y(y) \sim a+bx+\epsilon$. We chose $a=1$, $b=2$, $\sigma_\epsilon=0.5$ and $\Sigma_x=(-2, 2)$.

For the remainder of the ETAS parameters, we choose parameters based on the Tohoku District, Japan catalog from $1926-1995$ over $36^o$, $42^o$ N and $141^o$, $145^o$, E,  which were estimated using maximum likelihood by \cite{ogata1998space}. These are: $(\bar{K},\alpha,p,c,d,q) = (0.322,$ $1.407,$ $1.121,$ $0.0353,$ $0.0159,$ $1.531)$ with immigrant intensity constant $\tilde{\mu}= 0.854 \times 10^{-4}$ and margin of completion $M_0=5$.

The same parametrisation is used in \cite{fox2016spatially} subject to the following amendments -  $M_0=0$, $\Sigma=[0,4] \times [0,6]$,  and $\tilde{\mu} \phi(x_i,y_i)=0.001$ $+$ $0.004 \times \mathbbm{1}_{(x;y)} \{( [0,2]; [3,6] )$ $\cup$ $( [2,4]; [0,3] )\}  $ where $\mathbbm{1}_{(\cdot)} \{\cdot\}$ is an indicator function.  These simulations spread over temporal interval $[0, 25000]$. 

For our simulation, we choose to  set $\tilde{\mu}=0.325$ to provide denser catalogs within a shorter period of time. The overall event rate has increased by $0.35\times 10^4$ which allows us to run simulations for a shorter period of time compared to the previously introduced examples. However, this is not going to affect negatively the performance of the remaining parameters. 

All simulated catalogs in this section have magnitudes following the G-R law \citep{gutenberg1944frequency} with b-value of 1 i.e. $m_i-M_0 \sim Exponential\big[\beta=ln(10)\big]$. Hence, all marks $m$ are greater than the specific margin of completeness used for the simulation, $M_0$.  Within the simulation study, we set temporal window in $t \in (0, 300)$ with extension interval $\tau \in [300,350)$ and magnitude of completeness of $M_0=2$. 

\subsection{Model Fitting and Results}

For each of the three datasets, we used MCMC to draw $12,000$ samples from the posterior (after thinning). The branching structure was sampled from its conditional posterior only at every 50 iterations of the latent variable MCMC algorithm, since this is an $O(n^2)$ operation and slower than the other updates. For the DP ETAS model, we also resample the immigrant events density function $\phi(x,y)$ when a new branching structure is sampled. 

For the KDE ETAS model, the estimate of the immigrant spatial density $\phi(x,y)$ is based on the whole catalog of observations, and so is estimated prior to running the MCMC and set to a fixed value, as in \cite{zhuang2002stochastic, marsan2008extending, sornette2009limits, marsan2010new, fox2016spatially}.

In the simulation example we developed an out-of sample comparison with respect to 30 out-of-sample periods %(See Section \ref{section: extensions}) 
for each dataset. In order to evaluate the estimate performance we used every $50^{th}$ parameter set across the $10,000$ sets that were obtained as part of the MCMC procedure. This gives $200$ estimates of the out-of-sample log-likelihood for each of the $20$ catalog out-of-sample periods. Table \ref{tab:f123_summary} shows the average out-of-sample performance for the three models across these catalogs.

\begin{table}[ht!]
    \centering
    \begin{tabular}{|l||l|l|l|l||l|l|}
    \hline
    $\phi_\cdot$ & $DIC_{U}$ & $DIC_{K}$ & $DIC_{D}$ &   $\bar{l}^o_U$ &  $\bar{l}^o_K$ &   $\bar{l}^o_{D} $\\  
    \hline
$\phi_1$ &	5985.64 &	2495.93 & 4347.91 &   -1150.57 & 	-1079.11 & 	-1084.31 \\
$\phi_2$ &	5984.60 & 	2372.27 & 	4583.25 & 	-913.29 &	-862.19	& -832.30 \\
$\phi_3$ &	5918.93 &	1780.47 &	3670.92 &	-830.16 &	-777.72 &	-778.89 \\ 
\hline
    \end{tabular}
    \caption{Comparison between the performance of Unif (U), KDE (K) and DP (D) ETAS models across three uncaused events' spatial distributions ($\phi_\cdot$) with respect to the Tohoku District \citep{ogata1998space} MLE estimated based simulated catalogs. }
    \label{tab:f123_summary}
\end{table}

 The obtained results for $\phi_1(\cdot)$ and $\phi_3(\cdot)$ show that KDE ETAS model outperformed DP ETAS model, and that both outperformed the Fixed ETAS model (Table \ref{tab:f123_summary}). However, the results obtained based on $\phi_2(\cdot)$ show that DP is better than KDE with respect to all diagnostic tests. 

Since these results are mixed and show that both DP and KDE are capable of outperforming each other depending on the model parameters, we will now develop a larger simulation study to gain insight into the factors which determine when each is most suitable.

\subsection{Large Scale Simulation Study} \label{sec:multipleruns}
In order to examine further the behaviour of the spatial ETAS models, we created a number of simulated data sets by varying the model parameters. We set $\tilde{\mu}=0.325$, $c=0.0353$ and $p=1.121$ to be constant. Then we consider: $\alpha \in \{1.0, 1.3, 1.6, 1.9\}$, $\bar{K} \in \{0.1, 0.3, 0.5\}$, $d \in \{0.01, 0.255, 0.5\}$ and $d \in \{1.10, 1.55, 2\}$.  We exclude the combinations of parameters which result in an expected productivity greater than $1$ since these can potentially generate infinite catalogs as discussed in Section \ref{section:prior_choice_implementation_Details}. This resulted  in $63$ different parameter sets. 

Table \ref{tab:overal_summary_large_run} shows the results for DP and KDE ETAS on the 63 simulations, using the three specifications for $\phi(x,y)$ discussed in Section \ref{section:intro_f1_f2_f3} . This table presents the number of datasets that allocate either KDE or DP as the best model based on either DIC or out-of-sample maximum likelihood $\hat{l}^o$ or out-of-sample mean log-likelihood $\bar{l}^o$ or with respect to all previous metrics (referred as  \textit{best}). We further provided the aggregated counts across all 189 simulations.  It can be seen that both the DP and KDE versions of the model can outperform each other for different values of the model parameters. This shows that they are both likely to have value for estimating real-world catalogs.

It is interesting to understand the factors that make DP superior to KDE for particular data-sets, since this will give us a general rule for deciding which one is most appropriate to use. We propose the following hypothesis, which seems intuitively reasonable: since KDE forms its estimate of $\phi(x,y)$ by using all the earthquakes in the catalog rather than only the immigrant events, we would expect it to perform well either when most earthquakes are immigrants, or when the true distribution $\phi(x,y)$ of mainshocks is not too dissimilar to the overall distribution of earthquakes in the catalog.

We would expect the DP approach to perform better when $K$ is large (since this results in a higher proportion of aftershocks relative to immigrants), and also when the parameters $d$ and $q$ are large since this results in the distribution of triggered events being spread out over a wider areas, which increases the spatial discrepancy between the mainshock distribution and the overall catalog distribution.

\begin{table}[ht]
\begin{center}
\begin{tabular}{|c|c||c|c|c|c||c|}
    \hline
    subset & model & $max(\hat{l}_{\cdot})$ & $min(DIC)$ & $max(\hat{l}^o_{\cdot})$ & $max(\bar{l}^o_{\cdot})$ & best    \\
     \hline
      \hline
    \multirow{2}{*}{$\phi_1(\cdot)$}& KDE	& 49&	54&	49	&52&	47 \\
    &DP &14&	9&14&11&	8\\
\hline
        \multirow{2}{*}{$\phi_2(\cdot)$}&KDE	&26&	33&	25&	27&	23\\
    &DP&	37&	30&	38&	36&	29\\
 \hline
        \multirow{2}{*}{$\phi_3(\cdot)$}&KDE&	32&	35&	26&	27&	24	\\
&DP	&31&	28&	37&	36&	28\\
     \hline
      \hline
        \multirow{2}{*}{All}&KDE&	107&	122&	100&	106&	94\\
&DP	&82&	67&	89&	83	&65\\    
    \hline
\end{tabular}
\end{center}
 \caption{Number of datasets that allocate either KDE or DP as the best model based on either maximum log-likelihood $\hat{l}$ or DIC or out-of-sample maximum likelihood $\hat{l}^o$ or out-of-sample mean log-likelihood $\bar{l}^o$ or with respect to all previous metrics (\textit{best}). }
    \label{tab:overal_summary_large_run}
\end{table}

\begin{figure}[ht!]
\includegraphics[width=1\textwidth]{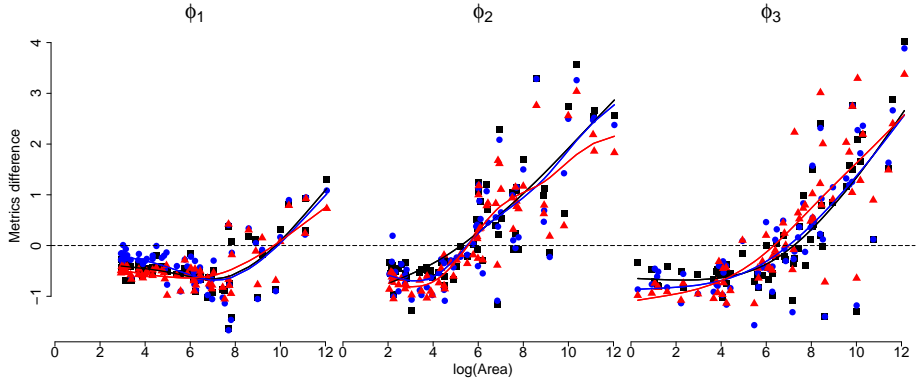}
\centering
\caption{Standardised differences of performance metrics of DP ETAS related to KDE ETAS with respect to the logarithmic transformation of  every catalog overall \textit{Area} across the three uncaused events' spatial densities $\phi_\cdot(\cdot)$ (Section \ref{section:intro_f1_f2_f3}). \textcolor{black}{\textrm{\ding{110}}}  stands for the difference between in-sample log-likelihood values for DP ETAS minus KDE ETAS;\textcolor{blue}{\textrm{ \ding{108}}}   stands for the difference between DIC values for KDE ETAS minus DP ETAS; \textcolor{red}{\textrm{\ding{115}}}  stands for the difference between out-of-sample log-likelihood values for DP ETAS minus KDE ETAS. For ease of display all values are re-scaled to follow a zero mean, unit variance Normal distribution. The three solid lines on each sub-plot represent the fitted lines of the pattern with respect to the three discussed difference (in their respective colours). The horizontal dashed line indicate the threshold for which DP ETAS will be considered to outperform KDE ETAS.}
\label{fig:DIc_ML_comparison}
\end{figure}

To test this hypothesis, we will try to create a single measure which represents the discrepancy between $\phi(x,y)$ and the overall catalog distribution.  This relationship is primarily influenced by the overall area that the catalog spans. Since all immigrant events are restricted to lie  within the same area ($\Sigma$), the overall area of the catalog is driven primarily by the values of $d$ and $q$ and $\bar{K}$, which affects how the triggered events spread out relative to the immigrants. As such, we compute the resulting areas for each of the $63$ catalogs which were simulated by varying the parameter values. In Figure \ref{fig:DIc_ML_comparison}, we plot how the area of the catalog relates to the degree to which DP performed KDE. Specifically, we plot the difference in the DIC and out-of-sample log-likelihood values between DP and KDE, as a function of catalog area. It can be seen that there is a clear relationship between the performance measures and the overall area of a catalog - a larger area is associated with a better performance of DP ETAS. The correlation between these measures and the proportion of immigrant events ($p_{\tilde{\mu}}$) is -0.27 and -0.26 for the DIC and ML differences respectively.  This largely confirms our previous hypothesis; in general, the DP model outperforms KDE when there are a large number of aftershocks that are spread out over a wide area, while KDE performs best when the number of aftershocks is smaller and more localised.

\section{Real earthquake sequences} \label{Section: Real_eartquakes}
In this Section we explore the performance of spatial ETAS model(s) across four different real earthquake catalogs. Since the Uniform model is unrealistic and performs substantially worse than the other two, we will not consider it further and instead focus on comparing the KDE and DP models.  A summary of the results is given in Table \ref{tab:summary_all_real_data_combined}.

\subsection{Vrancea, Romania}
Vrancea is an area in Romania that has a strong seismic influence on South-Eastern Europe. On $4/3/1977$ a large magnitude $7.2$ earthquake occurred which caused substantial destruction and human loss in both Bulgaria and Romania. We analysed the earthquake catalog from $01/01/1974$ to $01/01/2019$ that covers the spatial region $46^o$, $45^o 18'$ N and $27^o$, $26^o$ E with a magnitude of completeness of $M_0=2.5$. The data was split into a training set over the period $01/01/1975 - 01/01/2014$ with $529$ events and a test set over the period $01/01/2014 - 01/01/2019$ with comprises $46$ events. The data was obtained from the United States Geological Survey (USGS) catalog (\url{http://earthquake.usgs.gov/}).

The goodness-of-fit information is shown on Table \ref{tab:summary_all_real_data_combined}. It can be seen that DP ETAS  outperforms KDE when it comes to out-of-sample forecasting of the test set period. Interestingly, KDE does better when performance is evaluated using the DIC. Since this is computed using only the in-sample period, this suggests that KDE is overfitting to the data, while the DP method is estimating a more parsimonious model which results in superior forecasting. 

\subsection{Zakynthos and Kefalonia, Greece}
Zakynthos and Kefalonia are subject to prolonged seismic activity. The area of interest spans $38^o 33.54'$, $47^o 14.34'$ N and $21^o36.96'$, $19^o39.96'$ E.  The most important event in the region is the $6.8$ M Ionian earthquake that occurred on $12/08/1953$. Including data from this period is very challenging due to the high magnitude of catalog completeness caused by poor quality seismic detection equipment. For this reason we focused our study on a more recent time period from $(1/1/2069$ to $1/1/2019)$ and chose a magnitude of completeness of  $M_0=4.5$ to ensure consistency throughout the catalog. The data was split into a training set $01/01/1969 - 01/01/2018$ with $343$ events and a test set $01/01/2018 - 01/01/2019$ that comprises $109$ events. The data was obtained from the United States Geological Survey (USGS) catalog (\url{http://earthquake.usgs.gov/}). The results in Table \ref{tab:summary_all_real_data_combined} are the same as above: DP ETAS performs better for forecasting the test set, while KDE has a superior in-sample performance. Again, this suggests that KDE is overfitting and that the DP approach is more useful for forecasting.

\subsection{Friuli, Italy}
Friuli is an area in Italy that is primarily known for the $6.5$ M earthquake that occurred on $06/05/1976$, followed by multiple aftershocks with considerably large magnitudes. We based our study on the earthquake occurrence  from  $01/01/1975$ to $01/01/2019$ that covers the area within $46^o 36'$, $46^o$ N and $12^o18'$, $13^o30'$ E with minimum magnitude of $M_0=3$. For inferential purposes we split the data into a train set $01/01/1975 - 01/01/2004$ with $310$ events and a test set $01/01/2004 - 01/01/2019$ that comprises of $20$ events. The data was obtained from the United States Geological Survey (USGS) catalog (\url{http://earthquake.usgs.gov/}).

As before, the goodness-of-fit results are shown on Table \ref{tab:summary_all_real_data_combined} and show the same pattern as above: DP ETAS performs better when it comes to out-of-sample forecasting, due to KDE overfitting to the training set.

\subsection{Central Italy}
In 2006, Central Italy suffered a particularly damaging magnitude 6.2 earthquake that caused the death of nearly $300$ people \citep{luzi2017central}. Although the type of point process models discussed in this paper are not appropriate for predicting the occurrence of individual earthquakes, they are particularly useful for forecasting patterns in aftershock sequences. To this end, we obtained earthquake data from the Italian National Institute of Geophysics and Volcanology (Instituto Nazionale Di Geofisica e Vulcanalogia \url{http://www.ingv.it})  from $01/04/1999$ to $01/04/2019$ that spans Italy within $35^o$, $49^o$ N and $5^o$, $20^o$ E with magnitude of completeness taken to be $M_0=3$. Then we split the data into a training set from $01/04/1999$ to $01/04/2014$  ($4669$ events) which was used to estimate the model parameters, followed by a test set from $01/04/2014$ to $01/04/2019$ ($2171$ events) used to measure performance.

The model comparison results are shown on Table \ref{tab:summary_all_real_data_combined}. In this case, the KDE approach outperforms DP ETAS when it comes to out-of-sample forecasting. Investigating the catalog more closely, we found that most of the detected earthquakes were localised in a fairly small shore area. As discussed in the previous section, we would hence expect the KDE approach to perform particularly well here.

\begin{table}[ht!]
    \centering
 \begin{tabular}{|l||r|r||r|r|}
 \hline

       Data &   $DIC_{K}$ & $DIC_{D}$ &   $\bar{l}^o_K$ &   $\bar{l}^o_{D}$\\
       \hline
       \hline

%       Italy & 8882.07 & 8568.25 & - 6318.75 & -7403.23 &11586.14 & 11557.29 & 11580.85 & 11552.51 \\
        
%        Friuli  & 946.66 & 995.58 & -36.36 & -29.11  \\
 %       \hline
        Vrancea & 2710.23 & 2731.23  & -43.88 & -37.86\\
        \hline
        Zakynthos & 846.51 & 903.64  & -36.66 & -31.00\\
          \hline
        Friuli  & 946.66 & 995.58 & -36.36 & -29.11  \\
          \hline
       Italy  & - 6318.75 & -7403.23   & 11580.85 & 11552.51\\
          \hline
          
          \end{tabular}

    \caption{KDE and DP based spatial ETAS model comparison across real catalogs. Lower values of the DIC and larger (less negative) values of the out-of-sample likelihood indicate superior performance. The large value of the likelihood for the Central Italy catalog is due to the very large number of events compared to the other catalogs.}
    \label{tab:summary_all_real_data_combined}
\end{table}

\section{Conclusions} \label{Section:Conclusion}

The classic frequentist methods commonly used in seismic forecasting typically assume that model parameters are known exactly, which can result in forecasts which are overly confident. To mitigate this, Bayesian approaches are becoming more common in seismology. Despite being one of the most popular forecasting models, the ETAS framework has rarely received a fully Bayesian treatment, and even studies which attempt Bayesian forecasting often up resorting to frequentist-style plug-in estimates of the model parameters \citep{omi2015intermediate, ebrahimian2013adaptive}. This is due to the highly complex nature of the posterior distribution. In this work, we have introduced a new posterior sampling scheme which can be scaled up to catalogs containing thousands of earthquakes, and demonstrated its efficiency. Our approach can easily be deployed on realistic seismic catalogs containing thousands of earthquakes.

In this work, we explored the most commonly used version of the spatial ETAS model and showed how a nonparametric Dirichlet process prior could be used to allow fully Bayesian inference for the underlying spatial density. A comparison to a fixed Kernel Density Estimate of this density showed very promising performance in out-of-sample forecasting tasks. From our experiments on synthetic catalogs, we are able to draw the conclusion that the DP approach performs best when there are a large number of triggered events with a spatial distribution that is spread out over the region, while KDE is more suited to catalogs where all earthquakes occur in a compact area.

\begin{comment}
\section*{Acknowledgements}
\begin{itemize}
    \item Karen Kafadar, Editor-in-Chief, The Annals of Applied Statistics, for outlining critical aspect of our work related to its suitability for The Annals of Applied Statistics.  
    \item Miguel de Carvalho, Associate Editor, The Annals of Applied Statistics, for interesting suggestion for improvements of presentation of our methods. 
    \item Finn Lindgren, Associate Editor, The Annals of Applied Statistics
\end{itemize}

\end{comment}

\spacing{0.75}

\footnotesize

\bibliographystyle{apalike}
\bibliography{Reference}
\normalsize

\end{document}